\documentclass[showpacs,amsmath,amssymb,aps,pra,floatfix,nofootinbib,superscriptaddress]{revtex4}
\begin{document}

\title{ Comment on ``Dense coding in entangled states"}
\author{O. Akhavan}
\email{akhavan@mehr.sharif.edu}
 \address{Department of Physics,
Sharif University of Technology, P.O. Box 11365-9161, Tehran,
Iran}
\address{ Institute for Studies in Theoretical Physics and
Mathematics (IPM), P.O. Box 19395-5531, Tehran, Iran }
\author{A.T. Rezakhani}
\email{tayefehr@mehr.sharif.edu} \address{Department of Physics,
Sharif University of Technology, P.O. Box 11365-9161, Tehran,
Iran}
\begin{abstract}
In a recent Brief Report, Lee {\em et al.} [L. Lee, D. Ahn, and
S.W. Hwang, Phys. Rev. A {\bf{66}}, 024304 (2002)] have claimed
that using pairwise entangled qubits gives rise to an
exponentially more efficient dense coding when two parties are
involved than using maximally entangled qubits shared among $N$
parties. Here, we show that their claim is not true.
\end{abstract}
 \maketitle

Dense coding protocol as suggested by Bennett and Wiesner
{\cite{ref1}} is to send two bits of classical information by
sending just one qubit from the sender to the receiver which have
a shared entanglement. So, comparing with classical world where
there is no entanglement, one gets a more dense way of information
transfer by using quantum channels.

Recently, Lee {\em et al.} {\cite{ref2}} have considered two
different multiqubit schemes for dense coding and compared their
efficiency. In the pairwise scheme, Alice and Bob share $N$
separated maximally entangled pairs, so this scheme is equivalent
to $N$ separate dense coding schemes.  In their the maximally
entangled scheme, Alice has $N$ qubits and Bob possesses one,
which are prepared in a $N+1$ maximally entangled qubits. The
number of different unitary operators which can be constructed in
the first scheme is $4^N$ corresponding to the number of different
messages
 which can be sent. Since if we have $M$ different messages we can
codify them in ${\text{log}_{2} M}$ bits of information, then
${\text{log}_{2} 4^N}=2N$ bits of information can be transferred
by the pairwise scheme form Alice to Bob, by sending $N$
particles. Their essential mistake is just here by which they have
concluded that number of bits is $2^N$, which is incorrect.
Therefore the rates of information gain (bits per unit time) that
they have deduced in their Eqs. (6) and (7) for the pairwise and
maximally entangled schemes must change into

\begin{eqnarray}
&&r_p=\frac{2N}{N(t_h+t_c)}\nonumber\\
&&r_m=\frac{N+1}{t_h+Nt_c}
\end{eqnarray}
where $t_c$ and $t_h$ are operation times for CNOT and Hadamard
gates, respectively. If $t_c$ and $t_h$ are assumed equal it is
seen that, $r_p=r_m=\frac{1}{t_c}$, and there is no exponential
efficiency in the pairwise scheme. Therefore considering $N$
different Alices or combining them as a sole Alice (which changes
the number the parties involved but not the number of sent
particles) cannot lead into a more efficient protocol as they have
claimed. This is in accordance with the result of Bose {\em et
al.} {\cite{ref3}}.  As well, a more efficient (logarithmically)
has been proposed recently in \cite{ref4}.

Also, it should be noted that a more reasonable measure of
efficiency of the scheme can be the number of bits transferred in
the protocol per needed time per sent particles. By considering
this we have
\begin{eqnarray}
&&r_p= \frac{2}{N(t_h+t_c)}\nonumber\\
&&r_m=\frac{N+1}{N(t_h+Nt_c)}.
\end{eqnarray}
 Now these quantities give more better sense of efficiency of the
protocols, though, the number of the sent particles are the same
here. This concept of efficiency also applies to the scheme of
Bose {\em et al} {\cite{ref3}}.


\begin{thebibliography}{99}

\bibitem{ref1} C.H. Bennett and S.J. Wiesner, Phys. Rev. Lett.
{\bf{69}}, 2881 (1992).

\bibitem{ref2} L. Lee, D. Ahn, and S.W. Hwang, Phys. Rev. A {\bf{66}}, 024304 (2002).

\bibitem{ref3} S. Bose, V. Vedral, and P.L. Knight, Phys. Rev. A
{\bf{57}}, 822 (1998).

\bibitem{ref4} O. Akhavan, A.T. Rezakhani, and M. Golshani, Phys. Lett. A
{\bf{313}}, 261 (2003).
\end{thebibliography}
\end{document}